\documentclass[sigconf,letterpaper]{acmart}
\usepackage{balance}

\usepackage{xcolor}
\usepackage{enumitem}

\newcommand{\ie}{\emph{i.e., }}
\newcommand{\eg}{\emph{e.g., }}

\copyrightyear{2026}
\acmYear{2026}
\setcopyright{cc}
\setcctype{by}
\acmConference[CIKM '26]{Proceedings of the 35th ACM International Conference on Information and Knowledge Management}{November 07--11, 2026}{Rome, Italy}
\acmBooktitle{Proceedings of the 35th ACM International Conference on Information and Knowledge Management (CIKM '26), November 07--11, 2026, Rome, Italy}
\acmDOI{10.1145/3799682.3840096}
\acmISBN{979-8-4007-2539-5/2026/11}

\begin{document}

\title{Native Multimodal Representation Learning for Click-Through Rate Prediction in E-Commerce Scenarios}

\author{Chao Yi}
\authornote{Equal Contribution.}
\author{Feifan Yang}
\authornotemark[1]
\email{yunan.yc@alibaba-inc.com}
\email{yangfeifan.yff@alibaba-inc.com}
\affiliation{%
  \institution{Taobao \& Tmall Group of Alibaba}
  \city{Beijing}
  \country{China}
}

\author{Jiawei Feng}
\email{jwf3ng@gmail.com}
\affiliation{%
  \institution{University of Science and Technology of China}
  \city{Hefei, Anhui}
  \country{China}
}


\author{Sishuo Chen}
\author{Zhangming Chan}
\author{Xiang-Rong Sheng}
\author{Han Zhu}
\authornote{Corresponding Author.}
\email{chensishuo@pku.edu.cn}
\email{zhangming.czm@alibaba-inc.com}
\email{xiangrong.sxr@alibaba-inc.com}
\email{zhuhan.zh@alibaba-inc.com}
\affiliation{%
  \institution{Taobao \& Tmall Group of Alibaba}
  \city{Beijing}
  \country{China}
}

\renewcommand{\shortauthors}{Chao Yi, et al.}

\begin{abstract}
Multimodal representations have been widely adopted in industrial e-commerce recommendation systems. 
Due to their strong semantic understanding and generalization capabilities, they enhance the performance of traditional sparse ID-based Click-Through Rate
(CTR) prediction models. 
Current multimodal application frameworks in the CTR prediction task typically follow a \textbf{two-stage} paradigm: first, pre-training a multimodal encoder on data from specific recommendation scenarios; second, extracting items' multimodal representations using this pre-trained multimodal encoder and integrating them into the CTR prediction model.
However, the training objectives and data distribution of multimodal pre-training tasks often differ from those of the CTR prediction task, which limits the effectiveness of multimodal representation on downstream tasks.
In this paper, we focus on how to learn \textbf{Native Multimodal Representation} for the CTR prediction task.
One intuitive solution is to \textbf{jointly train} the multimodal encoder and CTR model end-to-end on the CTR task, with the expectation that the encoder can automatically learn downstream-relevant knowledge.
However, we find that the end-to-end training does not bring performance improvements to existing multimodal application paradigms.
Our analysis reveals that user behaviors in raw CTR data are driven by both multimodal semantics and non-multimodal factors, leading to ambiguous supervision and inconsistent encoder updates.
To address this, we propose a \textbf{Mine-Then-Train} method that mines high-quality, multimodally interpretable training samples from CTR data and uses them to fine-tune the multimodal encoder for better alignment with user click preferences.
Offline and online experiments demonstrate the effectiveness of our approach.
\end{abstract}




\maketitle

\section{Introduction}
Traditionally, industrial recommendation models have predominantly relied on discrete IDs~\cite{1_1,1_2,1_3,1_4,1_5,1_6,1_7,DIN}. However, ID-based models suffer from inherent limitations---notably, poor generalization in data-scarce scenarios~\cite{1_9,1_10,1_11,1_12} and limited ability to capture the rich semantics of items' multimodal content.

To address these limitations, recent work has attempted to incorporate the \textbf{multimodal representations} of items into ID-based models~\citep{simtier,qarm,onerec,notellm,notellm2,lemur,moon}. 
Typically, most approaches~\citep{simtier,qarm,onerec,notellm,notellm2,moon} adopt a \textbf{two-stage paradigm}, involving \textbf{1}) Pretraining a multimodal encoder, then \textbf{2}) Obtaining multimodal representations via this multimodal encoder and integrating them into the recommendation model. 
In the pretraining phase, existing methods collect multimodal training data from the corresponding business scenario and train multimodal encoders via contrastive learning~\citep{moco}, masked modeling~\citep{mae}, or generative tasks~\citep{coca} to endow them with the ability to understand and represent semantics in the specific business scenarios.
In the application phase, multimodal representations can be leveraged in multiple ways—such as serving as Semantic ID~\citep{qarm,onerec}, calculating the semantic similarity between items~\citep{simtier}, or being integrated directly as features in recommendation models.

However, since the pretraining phase of the multimodal model is unaware of the data and optimization objectives of downstream tasks~(\eg CTR prediction), a gap exists between the pretraining tasks and the actual requirements of the downstream tasks. 
This gap may limit the effectiveness ceiling of the multimodal representations in industrial recommendation systems.

\begin{figure*}[htbp!]
  \centering
  \includegraphics[width=0.95\linewidth]{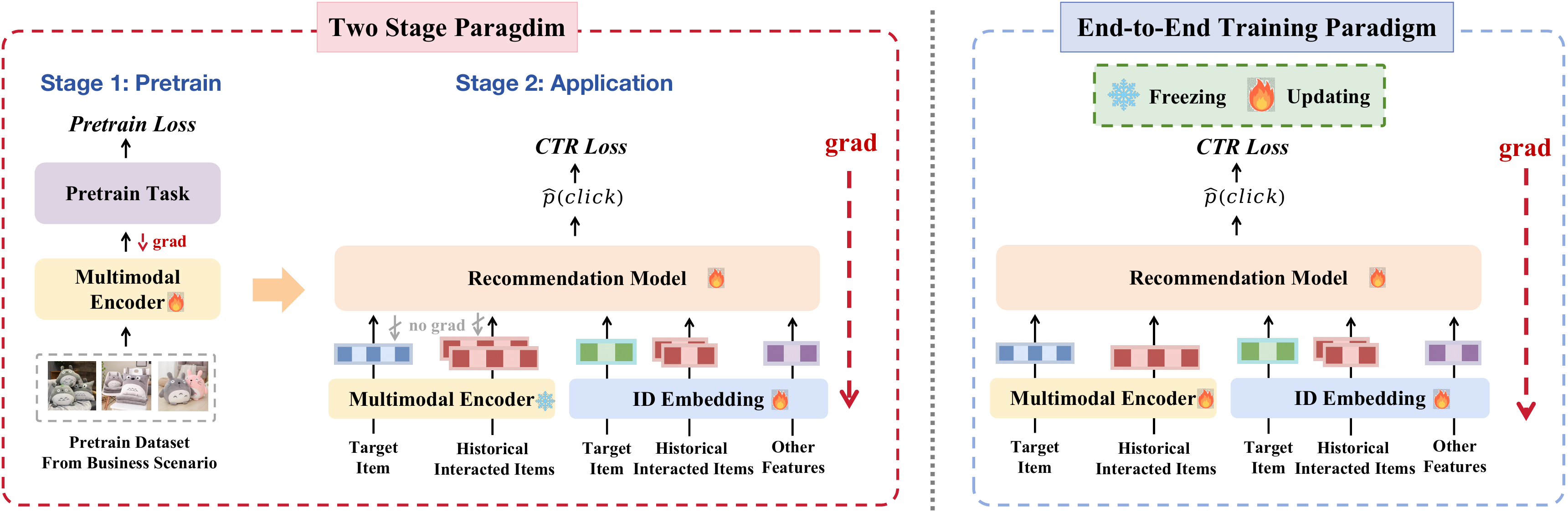}
  \vspace{-0.3cm}
  \caption{\textbf{Comparison between Two-Stage and End-to-End Training Paradigms.} The two-stage paradigm pretrains a multimodal encoder via tasks~(\eg MoCo~\cite{moco}) on business-specific data to capture scenario-specific semantics, then freezes it to extract items' representations for CTR prediction.  The end-to-end paradigm jointly optimizes the multimodal encoder and recommendation model directly on the CTR task, expecting to yield representations better aligned with CTR prediction requirements.}
    \label{fig1-1}
    \vspace{-0.3cm}
\end{figure*}

In this paper, we explore the \textbf{Native Multimodal Representation Learning for Click-Through
Rate Prediction}, with a primary focus on the E-Commerce Scenarios.
One natural way toward this goal is to adopt an \textbf{End-to-End Multimodal Training} paradigm---namely, integrating the multimodal encoder directly into the recommendation model and training the entire system end-to-end, which we refer to as $\textbf{E}$nd-to-$\textbf{E}$nd $\textbf{M}$ultimodal Training~(E2EM).
Figure \ref{fig1-1} illustrates the schematic diagrams.
We first seek to answer a question: \textbf{``Can $\textbf{E}$nd-to-$\textbf{E}$nd $\textbf{M}$ultimodal Training~(E2EM) yield additional performance gains on top of a high-quality, well-pretrained multimodal encoder for CTR prediction in E-Commerce Scenarios?''}
This question is particularly critical given that, in many practical industrial recommendation scenarios~\citep{simtier,qarm,onerec,notellm,notellm2,moon}, the two-stage paradigm has already yielded powerful multimodal encoders.
For instance, in Taobao's display advertising system, we adopt a multimodal encoder~(SCL Encoder~\citep{simtier}) pretrained on large-scale, high-quality e-commerce data.
Integrating the extracted multimodal representations led to a percentage-level improvement in GAUC. 
This represents a substantial improvement, as a 0.1\% GAUC lift increase is already recognized as a convincing result in such a large-scale system.
Our results indicate that E2EM \textbf{fails} to yield additional gains and even leads to degraded performance.

Motivated by these findings, we further analyze why E2EM fails.
We find that E2EM exposes the multimodal encoder to \textbf{ambiguous supervision from raw CTR data} for two reasons.
\textbf{First}, from the data perspective, user behaviors are only partially attributable to multimodal semantics and may instead reflect price comparison, accidental clicks, position bias, interest fatigue, and other non-semantic factors.
\textbf{Second}, from the optimization perspective, co-training with an ID-based CTR model cannot explicitly decouple these heterogeneous signals, because the CTR loss propagates each sample's gradient to the multimodal encoder.
Beyond these optimization challenges, E2EM also incurs substantial training costs.

To address this, we propose a \textbf{Mine-Then-Train} method that extracts multimodally interpretable supervision from CTR data before encoder training.
In the Mining stage, we train a multimodal annotation model on CTR data to capture multimodally interpretable click preferences.
This model is then utilized to automatically mine user-preference-aligned, and multimodally interpretable training samples from CTR data.
In the Training stage, we leverage the mined high-quality training data to fine-tune the pre-trained multimodal encoder, driving it to learn similarity ranking relations that better reflect user click preferences. 
By isolating supervision before encoder optimization, \textbf{Mine-Then-Train} avoids directly exposing the encoder to ambiguous CTR supervision and achieves significant performance gains in Taobao's display advertising system.

The main contributions are summarized as follows:
\begin{itemize}[leftmargin=*]
    \item We conduct $\textbf{E}$nd-to-$\textbf{E}$nd $\textbf{M}$ultimodal Training~(E2EM) experiments on a well-pretrained multimodal encoder in the display advertising scenario of Taobao—one of the world's largest e-commerce platforms—and the results show that E2EM does not yield performance gains.
    \item We show that user behaviors in CTR data are only partially attributable to multimodal semantics, providing ambiguous supervision and producing inconsistent encoder updates.
    \item We propose a \textbf{Mine-Then-Train} approach: first mine high-quality training data aligned with user click preferences from CTR data, then use it to train the multimodal encoder, enabling native multimodal representation learning tailored for CTR prediction. 
\end{itemize}
\section{Prelimilaries}
We first introduce the two-stage multimodal application paradigm for CTR prediction. 
In the first stage, a general multimodal encoder $E$ is adapted via pretraining tasks $\mathcal{T}_{\text{pretrain}}$ on collected scenario-specific training data $D_{\text{train}}$, yielding an encoder $\hat{E}$ with enhanced semantic representation capabilities tailored to the target scenario:
\begin{equation}
    \hat{E} = \mathcal{T}_{\text{pretrain}}(E, D_{\text{train}}).
\end{equation}
In the second stage, $\hat{E}$ extracts the multimodal representation from item's multimodal content~(such as image/title), which are incorporated into the CTR prediction model $F$:
\begin{equation}
    MMRep = \hat{E}(ItemMMInfo), \quad pCTR = F(IDFeat, MMRep).
\end{equation}
While these representations provide incremental gains to CTR prediction, they remain suboptimal due to two misalignments: 
(1) the pretraining task $\mathcal{T}_{\text{pretrain}}$ diverges from the CTR prediction task $\mathcal{T}_{\text{ctr}}$, and 
(2) the pretraining data $D_{\text{train}}$ differs in distribution from CTR data $D_{\text{ctr}}$. 
To address this, native multimodal learning paradigm further fine-tunes $\hat{E}$ based on the $D_{\text{ctr}}$ data:
\begin{equation}
    \tilde{E} = \mathcal{T}_{\text{ft}}(\hat{E}, D_{\text{ctr}}),
\end{equation}
It expects the encoder $\tilde{E}$ to yield representations better aligned with the CTR prediction task, leading to further improvements.
\section{Can E2EM Provide Gains over a Two-Stage Paradigm in the E-Commerce Scenario?}
\label{E2EM_failure}
In this section, we investigate \textbf{``Can E2EM provide gains on top of a strong two-stage paradigm in the E-Commerce Scenario?''}. 

\subsection{Implementation Details}
\label{util_method}
\noindent \textbf{Datasets.} 
We conduct the experiment on the dataset collected from Taobao, a well-known e-commerce platform.
The dataset contains click/non-click behavior data of 84M users over one week, which has 88M products and 1.9B samples.
Each sample is formatted as <user features, target item features, click label>. 
User features include age, gender, the sequence of items previously clicked by the user, and so on. 
Target item features include category, brand, and item ID, among others.
Moreover, each sample also includes multimodal content~(images, titles) for both the target item and each item in the user behavior sequence, which is used as input to the multimodal encoder to obtain multimodal representations.

\noindent \textbf{Models.} 
For the CTR prediction model, we employ the MUSE~\citep{MUSE} used in Taobao's display advertising system.
For the multimodal encoder, to reliably measure the value of E2EM, we adopt a well pre-trained multimodal encoder already proven effective in Taobao's display advertising system. 
Built on Chinese CLIP ViT-B/16~\citep{clip, chineseclip, vit}, it is pre-trained via \textbf{S}emantic-aware \textbf{C}ontrastive \textbf{L}earning~(SCL~\citep{simtier}), which applies MoCo-style~\citep{moco} contrastive learning on semantically similar pairs to capture fine-grained e-commerce semantics.

\noindent \textbf{Training Setting.} 
For the sparse embedding part of the CTR prediction model, we adopt the Adam~\citep{adam} optimizer with a learning rate of 2e-3. 
For the dense part of the recommendation model, we adopt the AdamW~\citep{adamw} optimizer with a learning rate of 2e-4. 
For the multimodal encoder, we also adopt the AdamW optimizer with a learning rate of 2e-4, along with a cosine learning rate scheduler. 
Further ablation studies are presented in Section \ref{more_ablation_study}. 

\noindent \textbf{Baseline.} 
We adopt the SCL~\citep{simtier} encoder deployed in Taobao’s display advertising system as our baseline.

\noindent \textbf{Multimodal Representations Usage Method.} 
We integrate multimodal representations into the CTR prediction model via two complementary ways. 
First, the Similarity-based method leverages multimodal representations to compute the similarity between the target item and each item in the user's historical click sequence.
This similarity information is incorporated into the CTR prediction model via methods like SA-TA~\citep{MUSE} and SimTier~\citep{simtier}.
Second, the Direct Fusion method merges each item's multimodal representation with its ID features via a linear layer, directly feeding the fused representations into the prediction model.

\noindent \textbf{Evaluation Metric.} 
We use GAUC~\citep{DIN,chen2025see,li2026delayed,luo2026modeling,wu2026mac}—a metric adopted in ranking tasks—to evaluate CTR prediction performance.  

\subsection{Results}
\begin{figure}[t]
\vspace{-0.3cm}
    \centering
    \includegraphics[width=\linewidth]{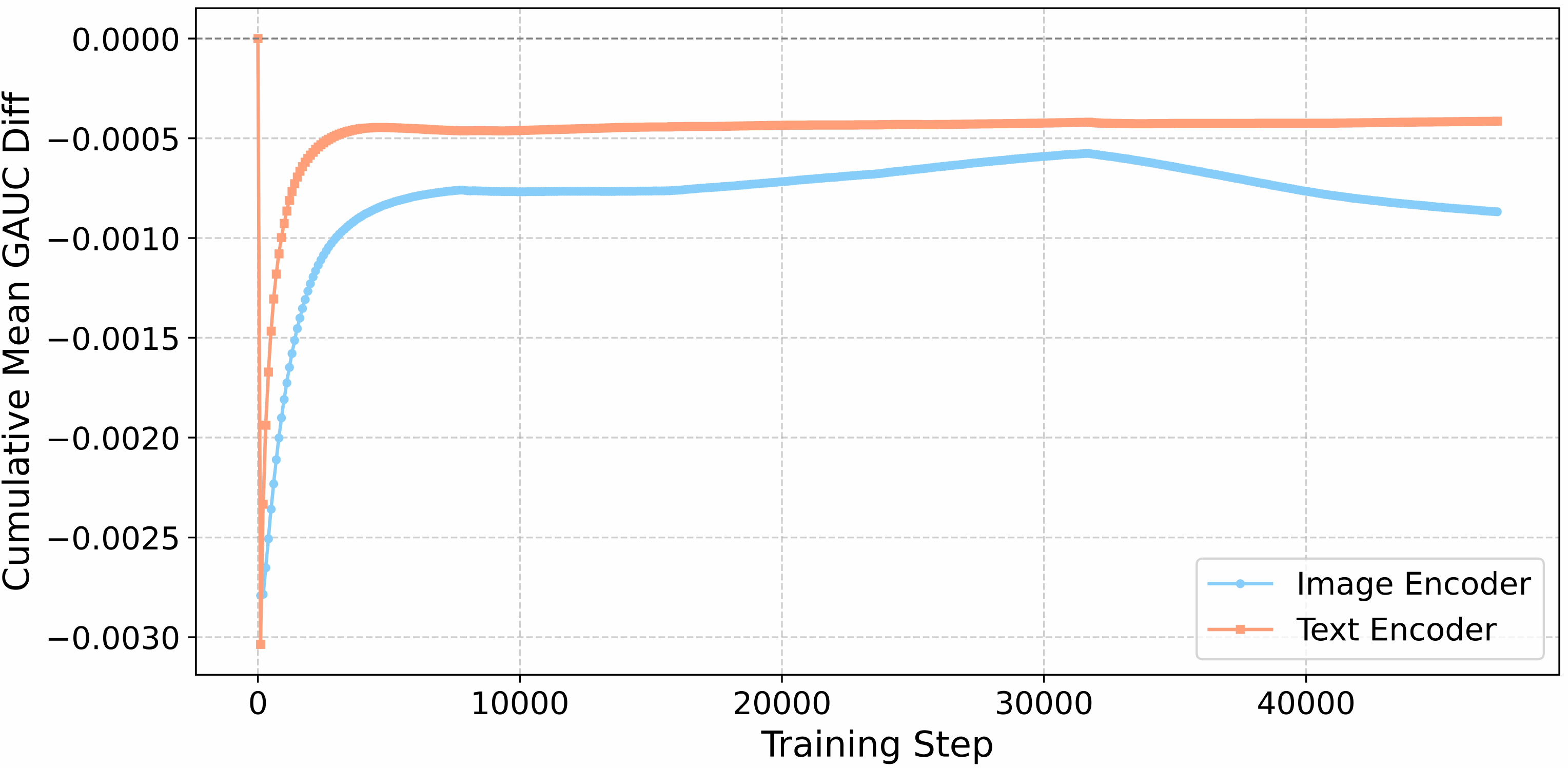}
    \vspace{-0.3cm}
    \caption{The GAUC difference curve comparing the CTR prediction model when using multimodal representations from the E2EM encoder versus those from the SCL encoder.}
    \vspace{-0.3cm}
    \label{fig.3-3}
\end{figure}

Figure \ref{fig.3-3} shows the GAUC difference curve comparing the CTR prediction model when using multimodal representations from the E2EM encoder versus those from the SCL encoder~\citep{simtier}.
From the figure we find that E2EM provides no improvement over the SCL encoder.
On the contrary, it leads to a drop in GAUC.

\subsection{More Ablation Study}
\label{more_ablation_study}
Due to the high training cost of E2EM in Taobao's real-world settings~(details in section \ref{compute_cost}), it is impractical to conduct multiple exploratory experiments applying E2EM.
To address this issue, we conduct experiments on Taobao-MM~\cite{MUSE}, a more lightweight benchmark. 
This benchmark collects data from Taobao's real-world scenarios and contains long user behavior sequences and multimodal representations, closely aligning with the settings of real industrial scenarios. 
We explore the following Research Questions~(RQ):
\begin{itemize}[leftmargin=*]
\item \textbf{RQ1: Scope of E2EM Application.} We evaluate E2EM under different application scopes: target-side only~(Tar E2EM), target-side combined with LEMUR's memory bank~\cite{lemur}~(Tar E2EM + Memory Bank), and E2EM on both sides~(Tar \& Seq E2EM). As shown in Table~\ref{table3-1}, all E2EM variants underperform the frozen-encoder baseline~(MUSE), with performance consistently degrading as the scope of E2EM expands. 
The reason is that E2EM representations underperform SCL representations, expanding their application scope merely exacerbates the performance decline.

\begin{table}[htbp!]
    \centering
    \caption{Experiment result of RQ1. MUSE denotes that both the target and sequence sides use SCL representations.}
    \vspace{-0.3cm}
    \label{table3-1}
    \begin{tabular}{c|c}
    \toprule
  E2EM Method & GAUC \\
  \midrule
  MUSE~\citep{MUSE} & \textbf{0.6154} \\
  Tar E2EM & 0.6139 \\
  Tar E2EM + Memory Bank~\citep{lemur} & 0.6128 \\
  Tar \& Seq E2EM & 0.6124 \\
  \bottomrule
    \end{tabular}
\end{table}
\begin{table}[t]
    \centering
    \caption{Experiment result of RQ2.}
    \vspace{-0.3cm}
    \label{table3-2}
    \begin{tabular}{c|c|c|c}
    \toprule
  GAUC & lr 2e-3 & lr 2e-4 & lr 2e-5 \\
  \midrule
  SGD & 0.6139 & 0.6148 & 0.6146 \\
  Adam & 0.6135 & 0.6133 & \text{0.6140} \\
  AdamW & 0.6134 & \text{0.6139} & \textbf{0.6151} \\
  \bottomrule
    \end{tabular}
    \vspace{-0.3cm}
\end{table}

    \item \textbf{RQ2: Learning Rate \& Optimizer Choice.} We replace the optimizer of the multimodal encoder with SGD, Adam, and AdamW, respectively, and explore its learning rates of 2e-3, 2e-4, and 2e-5, to eliminate the influence of training hyperparameters on the conclusions. Table \ref{table3-2} shows that under all settings, E2EM consistently underperforms the MUSE Baseline~(GAUC 0.6154).
\end{itemize}

\section{Why Does E2EM Fail with a Strong Encoder in the E-Commerce Scenario?}
In this section, we analyze why E2EM fails to improve a multimodal encoder with strong semantic understanding in the e-commerce scenario. 
We ask whether raw CTR data provide reliable multimodal supervision and whether co-training with an ID-based CTR model prevents non-multimodal factors from disrupting the encoder.

\subsection{E-Commerce CTR Data Are Only Partially Attributable to Multimodal Semantics}
\label{data_perspective}
\begin{figure}[htbp!]
\vspace{-0.3cm}
    \centering
    \includegraphics[width=0.8\linewidth]{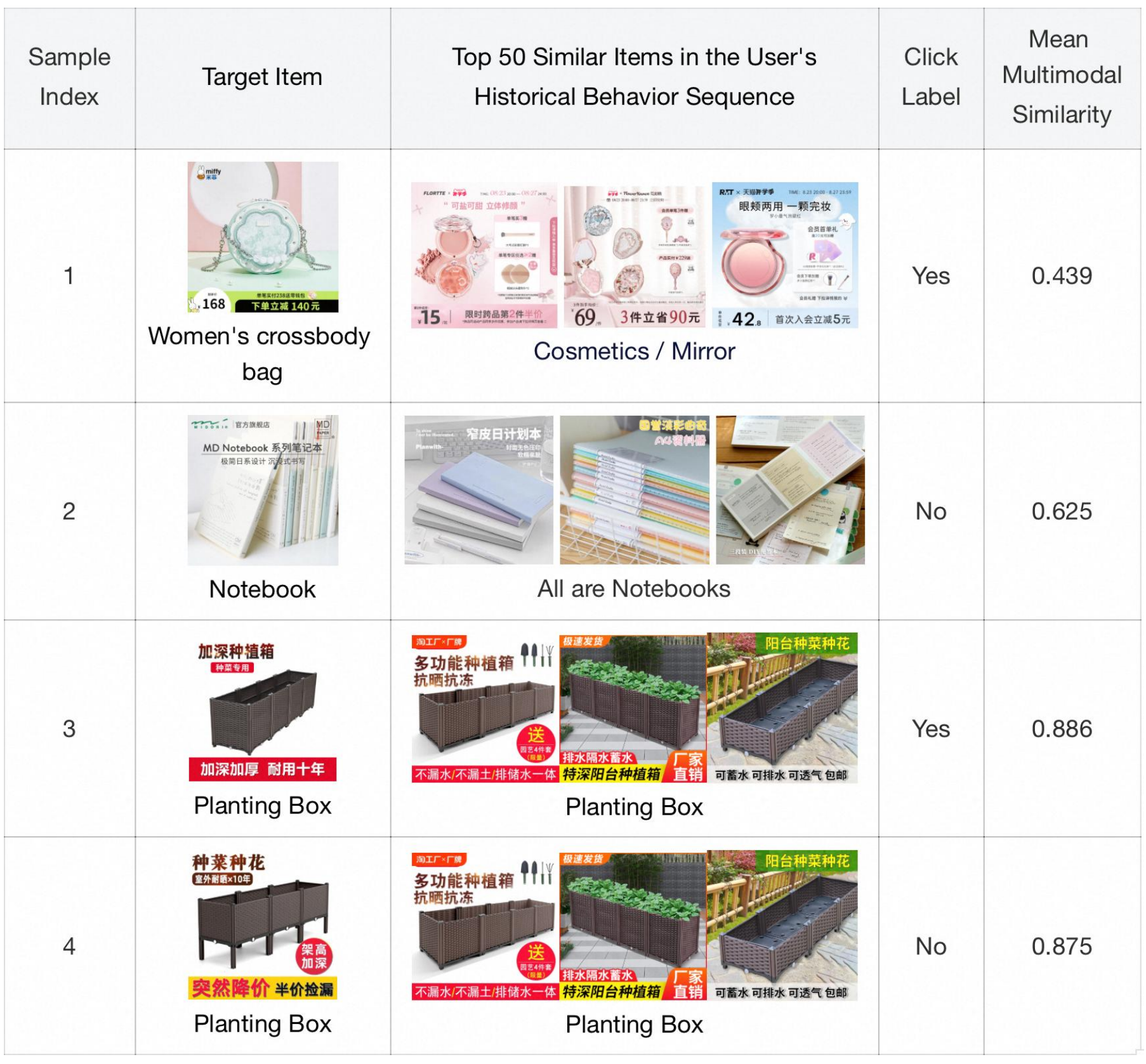}
    \vspace{-0.3cm}
    \caption{Examples of CTR data that are difficult to explain from the multimodal perspective.}
    \label{fig.4-1}
    \vspace{-0.3cm}
\end{figure}

We first examine to what extent e-commerce CTR data can be attributed to multimodal content.
Following MUSE~\citep{MUSE}, we use SCL representations to retrieve the Top-K semantically similar items from each user's behavior sequence and visualize them with the target item. 
Figure~\ref{fig.4-1} reveals two representative mismatches between content semantics and clicks:
\begin{itemize}[leftmargin=*,nosep]
  \item Price-comparison clicks, accidental clicks, position bias, and other non-semantic factors may cause users to click low-similarity items while skipping highly similar ones~(Samples 1\&2).
  \item Interest fatigue may cause users to skip content similar to a previously clicked item~(Samples 3\&4).
\end{itemize}

\begin{figure}[t]
    \centering
    \includegraphics[width=0.6\linewidth]{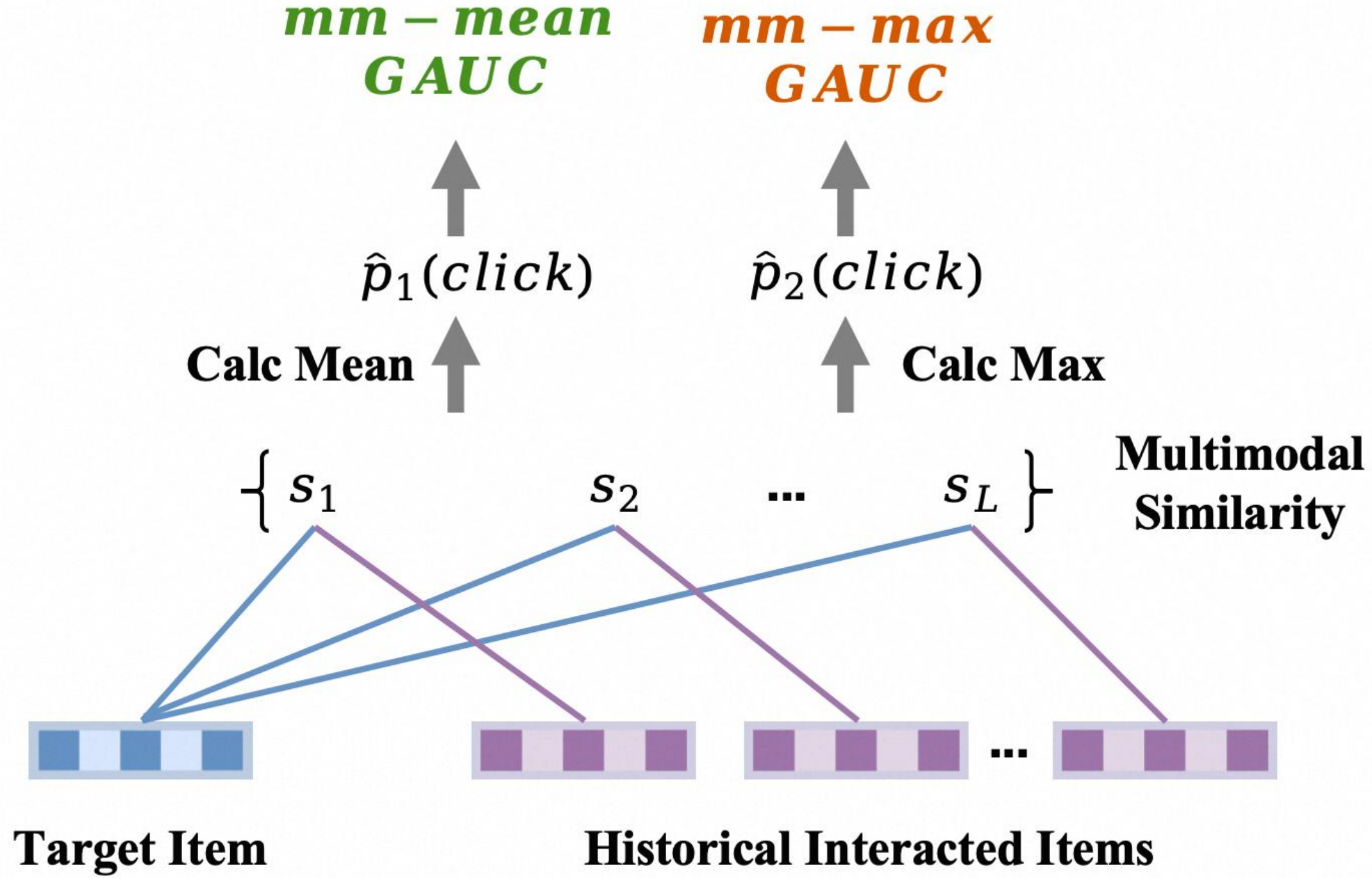}
    \vspace{-0.3cm}
    \caption{Computation of MultiModal-Mean GAUC and MultiModal-Max GAUC.}
    \label{fig.4-2}
    \vspace{-0.3cm}
\end{figure}

We further quantify this mismatch using two metrics. As illustrated in Figure~\ref{fig.4-2}, we compute the similarities between the target item and the user's historically clicked items using SCL representations, and use their mean and maximum as prediction scores. The resulting MultiModal-Mean GAUC and MultiModal-Max GAUC measure how well multimodal semantics similarity alone aligns with click preferences in the e-commerce setting. 
As shown in Table~\ref{table4-1}, both metrics are only slightly above random prediction~(0.5), showing that multimodal similarity alone is insufficient to explain a substantial portion of CTR data. 
This does not imply that CTR data contain no useful multimodal signals. 
Rather, it shows the CTR labels mix multimodal and non-multimodal factors, making the raw CTR data not a pure multimodal supervision signal.
Consequently, raw CTR data provide ambiguous supervision for directly optimizing a shared multimodal encoder.
\begin{table}[h]
    \vspace{-0.3cm}
    \centering
    \caption{Comparison between MultiModal-Mean GAUC and MultiModal-Max GAUC.}
    \vspace{-0.3cm}
    \label{table4-1}
    \begin{tabular}{c|c}
    \toprule
  MM Mean GAUC~($\uparrow$) & MM Max GAUC~($\uparrow$) \\
  \midrule
  0.541 & 0.536 \\
  \bottomrule
    \end{tabular}
  \vspace{-0.3cm}
\end{table}

\subsection{Co-Training Does Not Automatically Decouple CTR Signals}
The observations above suggest a solution: co-train the encoder with the ID-based CTR model, with the expectation that the ID branch absorbs non-multimodal factors while the encoder learns signals attributable to multimodal semantics.
Yet Section~\ref{E2EM_failure} shows that this expected division does not emerge.
The presence of the ID-based CTR model branch provides additional modeling capacity, but it does not explicitly assign each supervision signal to its responsible branch.
We further analyze this failure through the gradients produced by E2EM co-training.
Specifically, while keeping the encoder parameters frozen to control for variables, we measure the gradients with respect to the multimodal encoder's last Transformer layer across varying input data batches during end-to-end co-training with the ID-based CTR model. 
We then compare these gradients with those of SCL~\citep{simtier}.
\textbf{Avg Sim 1} is the mean cosine similarity between sample gradients within a batch, while \textbf{Avg Sim 2} is the cosine similarity between mean gradients of adjacent batches; lower values indicate less consistent optimization directions.

\begin{table}[h]
    \centering
    \vspace{-0.3cm}
    \caption{Gradient alignment result for different tasks.}
    \vspace{-0.3cm}
    \label{table4-2}
    \begin{tabular}{c|c|c}
    \toprule
  Task & Avg Sim 1~($\uparrow$) & Avg Sim 2~($\uparrow$) \\
  \midrule
  SCL~\cite{simtier} & 0.144 & 0.849 \\
  E2EM~(Co-train CTR) & 0.016 & 0.006 \\
  \bottomrule
    \end{tabular}
  \vspace{-0.3cm}
\end{table}

Table~\ref{table4-2} shows that E2EM has substantially lower gradient similarities than SCL despite the presence of the ID branch, indicating less consistent encoder updates. 
Together with the performance degradation in Section~3, this result suggests that CTR supervision is not effectively routed between the two branches.
The shared CTR loss helps explain this result: it backpropagates the loss gradient from each CTR sample through both branches without determining whether the current CTR data is explainable by multimodal content. 
Consequently, although the ID-based CTR model may fit such behavior, it alone cannot prevent supervision signals caused by non-multimodal factors from updating the multimodal encoder.

We further add SCL Loss~\citep{simtier} to anchor the encoder to multimodal semantics. If insufficient semantic constraint causes the interference, SCL should stabilize training. However, after introducing SCL Loss, CTR GAUC drops by \textbf{0.17} percentage points~(compared to not introducing SCL loss), while SCL Top-1 Accuracy decreases by \textbf{5} pt~(compared to the original SCL Encoder).
Although SCL supplies semantic supervision, it still cannot identify which CTR behaviors should update the encoder. 
These results motivate our \textbf{``Mine-Then-Train''} strategy, which isolates supervision at the data level so that only reliable, multimodally interpretable signals update the encoder, rather than relying on additional joint regularization.

\subsection{Substantial Training Cost}
\label{compute_cost}
Independent of the optimization issue above, another bottleneck of E2EM is its prohibitive training cost, caused by the forward and backward propagation over massive multimodal inputs.
With batch size $B$ and sequence length $N$, applying E2EM to both target and historical items processes $B \times (N+1)$ images/texts per step. We measure peak GPU memory and average step time using CLIP ViT-B/16 and MUSE~\citep{MUSE}~($B=32$, $N=50$). 
Both GPU memory and step time increase substantially, as shown in Table~\ref{table4-3}.
This cost will become increasingly unacceptable as industry adopts representation models with ever-larger parameter counts~\citep{notellm,notellm2,moon}.

\vspace{-0.3cm}
\begin{table}[htbp!]
    \centering
    \caption{Computation Cost of E2EM.}
    \vspace{-0.3cm}
    \label{table4-3}
    \setlength{\tabcolsep}{3pt}
    \begin{tabular}{c|c|c}
    \toprule
  Metric & Peak GPU Mem~(GB) & Avg Step Time~(s)\\
  \midrule
  Without E2EM & 3.37 & 1.02 \\
  With E2EM~(tar) & 5.71~(\textcolor{red}{\textbf{1.69$\times$}}) & 1.94~(\textcolor{red}{\textbf{1.90$\times$}}) \\
  With E2EM~(tar\&seq) & 84.56~(\textcolor{red}{\textbf{25.09$\times$}}) & 7.32~(\textcolor{red}{\textbf{7.18$\times$}}) \\
  \bottomrule
    \end{tabular}
    \vspace{-0.3cm}
\end{table}

Overall, direct E2EM co-training is both difficult to optimize and computationally expensive, motivating the decoupled \textbf{``Mine-Then-Train''} framework introduced in Section~5.
\section{Mine-Then-Train: Native Multimodal Repre-\\-sentation Learning for CTR Prediction}
In this section, we introduce a \textbf{``Mine-Then-Train''} method to learn native multimodal representations for CTR prediction.

\subsection{Why Do Noisy Samples Have a Significant Impact on Multimodal Representations?}
\label{mm_vs_id}
To understand why multimodal representations are more sensitive to noise than ID features, we highlight their key difference: whether parameters are shared during item representation extraction.

\begin{itemize}[leftmargin=*]
  \item ID features assign each item a dedicated set of learnable parameters, resulting in a highly decoupled learning process where items largely do not influence one another.
  \item In contrast, multimodal encoders extract representations for all items using a single, fully shared multimodal encoder, thereby introducing strong parameter coupling across items.
\end{itemize}
While parameter sharing in multimodal representation extraction improves generalization, it introduces a critical problem: gradients from noisy samples~(\ie samples not attributable to multimodal semantics) can corrupt the shared encoder parameters, degrading representation quality across all items. 
This renders multimodal encoders highly vulnerable to CTR data's noise, whereas ID features are inherently robust due to their decoupled parameters. Rather than relying on additional joint regularization to suppress such noise, we extracts multimodally interpretable supervision from CTR data and then use only the mined data to train the encoder.
To this end, we propose a \textbf{``Mine-Then-Train''} approach.

\subsection{Automatically Mining Training Samples from CTR Data via Annotation Model}
\label{section-5_2}
A straightforward approach is to fine-tune the multimodal encoder on item pairs with high collaborative similarity~(\eg high Swing~\citep{swing} scores or co-click counts). 
However, our experiments show that this yields no noticeable gain over our SCL encoder~\citep{MUSE}.
We attribute this to two reasons: \textbf{(1)~Information Redundancy:} Collaborative signals in these pairs are already well captured by ID features in the CTR model, making the learned representations highly redundant with existing ID features and offering limited information gain; \textbf{(2)~Semantic Inconsistency:} Some pairs lack clear multimodal semantic relations~(\eg visual similarity), which may degrade the fine-grained multimodal semantic understanding ability of the pretrained SCL encoder.
To address these issues, we propose a selective data mining strategy that extracts training samples from CTR data satisfying three criteria:
\textbf{(1)~Aligning with user click preferences, (2)~Explainable from the multimodal semantic perspective, and (3)~Offering information gain beyond existing multimodal representations and ID features.}

\begin{figure}[t]
    \vspace{-0.3cm}
    \centering
    \includegraphics[width=0.76\linewidth]{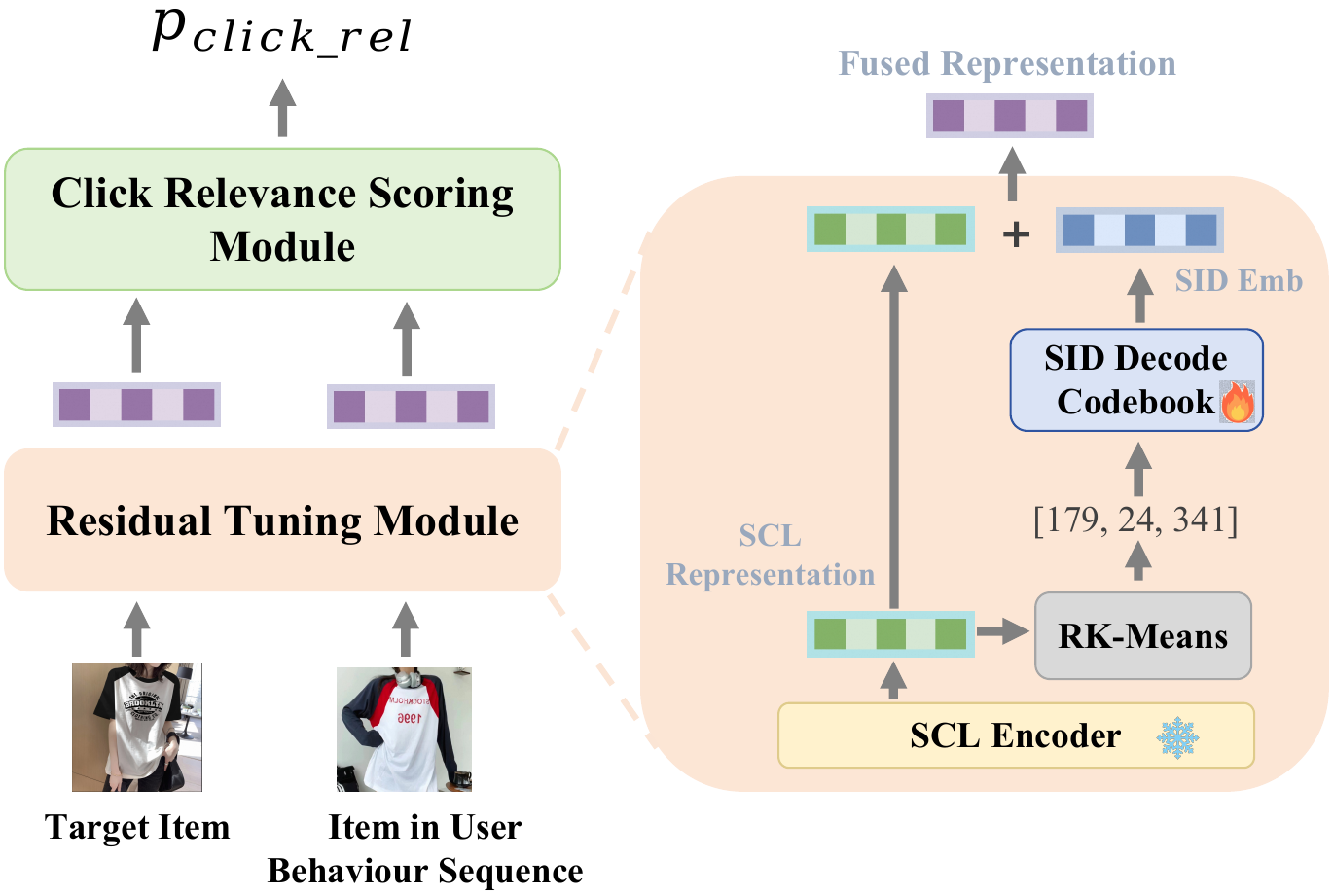}
    \vspace{-0.3cm}
    \caption{Architecture of the Multimodal Annotation Model.}
    \label{fig.5-1}
    \vspace{-0.3cm}
\end{figure}

Our data mining process consists of two steps: 
\textbf{First}, we train a multimodal annotation model on CTR data to capture user click preferences explainable by multimodal semantics. \textbf{Second}, we use its outputs to mine high-quality multimodal training samples from the CTR data.
In the first stage,given the target item and a user behavior item retrieved by GSU~\citep{simtier} using SCL representations, the annotation model uses their raw multimodal content~(\eg main images and titles) to predict a click-relevance score.
This score is integrated into the downstream CTR prediction model through methods such as SA-TA~\citep{MUSE} and SimTier~\citep{simtier}, allowing the CTR loss to train the annotation model end-to-end.
To preserve the SCL encoder's fine-grained understanding of e-commerce semantics, we adopt the residual-learning design of TaskRes~\citep{taskres} for the annotation model, whose architecture is shown in Figure~\ref{fig.5-1}.
In the \textbf{Residual Tuning Module}, the frozen SCL encoder first extracts representations from each item's raw multimodal content.
Then, RK-Means~\citep{qarm} clustering over these representations assigns each item an SID.
The corresponding SID Embeddings are retrieved from a learnable $3 \times 8192$ SID Decode Codebook and serve as CTR-specific residuals to the SCL representation.
The decode codebook is zero-initialized so that training begins without perturbing the original SCL representations.
The resulting representations of the two items are then passed to the \textbf{Click Relevance Scoring Module}, which computes the click-relevance score using either a Multi-Layer Perceptron~(MLP) or a normalized inner product.
We highlight the key architectural considerations of the annotation model:
\begin{itemize}[leftmargin=*]
  \item \textbf{Why use Residual Tuning}: In Residual Tuning, user preferences are captured entirely by the SID Decode Codebook without altering the SCL Encoder, thereby preventing non-multimodal factors in CTR data from degrading the encoder's fine-grained e-commerce semantic understanding capability. Furthermore, this design enables the codebook to explicitly learn the incremental information on top of SCL representations for the CTR task.
  \item \textbf{Why use SID Embeddings}: Item SIDs are closely tied to item semantics, ensuring that the learned information is multimodally interpretable. Additionally, tuning SID Embeddings is much more efficient than tuning the multimodal encoder, as SIDs can be pre-computed and stored beforehand. Thus, the end-to-end training phase only requires a simple embedding lookup from the decode codebook. Finally, the decoupled parameters of SID Embeddings enable more effective learning from CTR data.
\end{itemize}

\begin{figure}[h]
    \vspace{-0.3cm}
    \centering
    \includegraphics[width=0.76\linewidth]{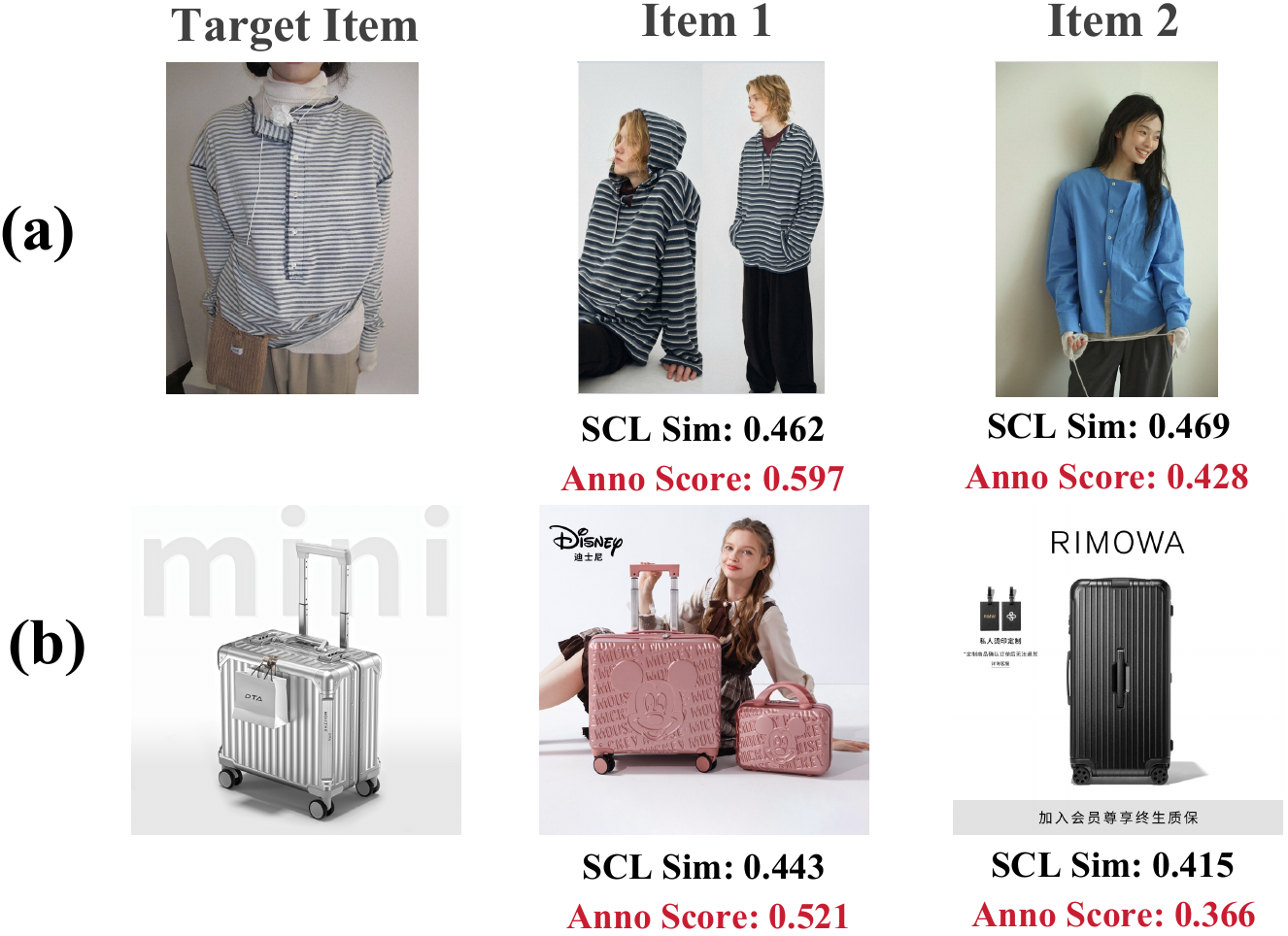}
    \vspace{-0.3cm}
    \caption{Examples comparing similarities of SCL representations~(SCL Sim) with annotation model scores~(Anno Score).}
    \label{fig.5-2}
\end{figure}

Integrating the multimodal annotation model's scores as features into the CTR prediction model yields a 0.13\% GAUC lift, which demonstrates that the annotation model indeed learns effective informational gain.
To investigate what the annotation model has learned, we compare the cosine similarities of item pairs calculated from SCL representations with the click relevance scores predicted by the annotation model.
Figure \ref{fig.5-2} presents several examples. 

\vspace{-0.3cm}

The annotation model learns to \textbf{focus on item attributes that more strongly drive user click preferences}. 
In case (a), despite the target item and item 2 sharing more similar silhouette designs~(hoodless with the same neckline design), the annotation model identifies matching stripe patterns as the more important factor driving user clicks. 
In case (b), although the target item and item 2 have more similar material textures~(vertical grooves), the model recognizes luggage size as the more decisive factor.

While capturing incremental information, the annotation model still faces two limitations: 
(1) since it is trained directly on CTR data, non-multimodal factors can make its click-relevance scores inaccurate for some item pairs;
and (2) user preferences are learned solely through SID embeddings, whose learning process essentially resembles the memorization mechanism of ID embeddings and lacks direct access to the raw multimodal content of items, limiting the generalization of the learned knowledge. 
To address this, we filter high-quality, clean labels produced by the annotation model to construct multimodal fine-tuning data for the SCL Encoder.

Specifically, for each CTR data sample, we use SCL representations to retrieve the Top-K similar items to the target item from the user's click history, and then pair each of these K items with the target item, filtering out item pairs with multimodal similarity below $\tau_m$ to ensure strong semantic relevance within each pair.
We then combine pairs sharing the same target item into triplets, and retain those satisfying the following conditions:
\begin{equation}
\begin{gathered}
    Sim_{\text{tar},1}^{\text{SCL}}-Sim_{\text{tar},2}^{\text{SCL}} < \tau_s, \ 
    Score_{\text{tar},1}^{\text{Anno}}-Score_{\text{tar},2}^{\text{Anno}} > \tau_a
\end{gathered}
\label{eq.5}
\end{equation}
Here, $Sim_{\text{tar},1}^{\text{SCL}}$ and $Sim_{\text{tar},2}^{\text{SCL}}$ denote the SCL-based similarities between the target item and the other two items in the triplet, while $Score_{\text{tar},1}^{\text{Anno}}$ and $Score_{\text{tar},2}^{\text{Anno}}$ represent their corresponding relevance scores from the annotation model.
$\tau_m$, $\tau_s$, and $\tau_a$ are the corresponding thresholds for pair similarity, SCL similarity gap, and annotation-score margin.
Together, these constraints retain semantically related triplets with clear preference differences not captured by SCL. In particular, the annotation-score margin measures the annotation model's confidence in the predicted preference ranking.

To verify whether this margin reliably indicates label quality, we conduct a human evaluation experiment.
Specifically, with $Sim_{\text{tar},1}^{\text{SCL}} - Sim_{\text{tar},2}^{\text{SCL}} < 0.05$, we sample 1000 high-confidence triplets with an annotation-score margin above 0.15 and 1000 low-confidence triplets with a margin below 0.05.
 Multiple human annotators assess whether a user is more likely to click Item 1 or Item 2 after clicking the target item through multi-round voting, with the majority vote serving as ground truth.
 Item 1 achieves a win rate of \textbf{85\%} in the high-confidence group, compared with only \textbf{42\%} in the low-confidence group. This substantial gap confirms that the annotation-score margin is a reliable indicator of pseudo-label quality and can effectively filter noisy triplets.

The retained data include two types of triplets: \textbf{(1) Rank-reversed triplets}, where Item 2 is initially more similar to the Target Item than Item 1 under the SCL representations, but the ranking is clearly reversed under the annotation model scores; and \textbf{(2) Gap-widened triplets}, where Item 1 remains more similar to the Target than Item 2, but the annotation model yields a significantly wider score gap.
Both provide fine-grained supervision that directs the encoder to focus on item attributes that most influence user click preferences.
Finally, we get a dataset of 30 million high-quality triplets.

\subsection{Training Native Multimodal Encoder for CTR Prediction Based on Mined Data}
\label{section-5_3}
Based on the triplets mined in Section~\ref{section-5_2}, we fine-tune the SCL encoder using the triplet margin loss to learn similarity rankings:
\begin{equation}
    L_{\text{margin}} = \max\left(0, m - (Sim_{\text{tar},1} - Sim_{\text{tar},2})\right),
    \label{eq.4}
\end{equation}
where $Sim_{\text{tar},1}$ and $Sim_{\text{tar},2}$ represent the cosine similarities calculated from representations extracted by the multimodal encoder currently being fine-tuned, and $m$ is the margin threshold set to 0.1.
To prevent the SCL encoder from forgetting the fine-grained e-commerce semantic understanding capabilities learned during its SCL pre-training phase, we compute the SCL Loss~\citep{simtier} and incorporate it as a regularization term in the model training process:
\begin{equation}
    L_{SCL} = -log\frac{\text{exp}(q\cdot k_{+}/\tau)}{\sum_{i=0}^{K}\text{exp}(q\cdot k_i/\tau)}.
    \label{eq.5}
\end{equation}
Here $q$ and $k_{+}$ denote the user's queried image/text and its corresponding purchased item, respectively, while $k_0, k_1, \dots, k_{K}$ represent other negative items stored in the memory queue, where $K$ denotes the memory queue size in MoCo~\cite{moco}.
$\tau$ represents a learnable temperature parameter. 
All items' representations in Equation \ref{eq.5} are L2-normalized.
The final total loss is given as follows:
\begin{equation}
    L_{total} = L_{margin} + L_{SCL}.
    \label{eq.6}
\end{equation}
This loss function enables the multimodal encoder to extract native representations aligned with user click preferences.

\subsection{Online System Deployment Pipeline}
Since this work only optimizes the multimodal encoder without modifying any other modules, we can directly reuse the SCL~\citep{simtier} encoder's online system deployment pipeline.

\section{Experiments}
In this section, we conduct experiments on the Taobao display advertising system to validate the effectiveness of the proposed \textbf{N}ative \textbf{M}ultimodal \textbf{R}epresentation \textbf{L}earning~(NMRL) method.

\subsection{Implementation Details}
\noindent \textbf{Dataset.} The CTR prediction dataset is obtained from the Taobao display advertising system, covering a one-month period.

\noindent \textbf{Model.} We adopt the CTR prediction model~\citep{est} from Taobao's display advertising system, which has already incorporated strong SCL multimodal representations, making it a strong baseline.

\noindent \textbf{Mining Thresholds.} We set the pairwise multimodal-similarity threshold $\tau_m$ to 0.3, the SCL similarity-gap threshold $\tau_s$ to 0.05, and the annotation score-margin threshold $\tau_a$ to 0.15.

\noindent \textbf{Multimodal Representation Usage Method.} 
We incorporate the \textbf{N}ative \textbf{M}ultimodal \textbf{R}epresentation~(NMR) into the sequential modeling module of the CTR prediction model in two ways: (1) Computing the similarity between the target item and the user's historical click sequence using NMR, and passing it to subsequent modeling modules~(SimTier~\citep{simtier}, SA-TA~\citep{MUSE}); (2) Directly fusing the NMR with other ID embeddings of the item via a linear layer, and passing the fused representations into the subsequent module.

\noindent \textbf{Evaluation Metric.} We use AUC and GAUC~\citep{DIN,chen2025see,li2026delayed,luo2026modeling,wu2026mac} to evaluate the CTR prediction model's ranking performance.

\subsection{Results \& Analysis}
Table \ref{table6-1} shows the relative offline improvement in AUC and GAUC over the baseline.
Here, \textbf{``Add Annotation Score''} denotes integrating the annotation model's scores into the CTR prediction model.
\textbf{``NMRL''} refers to fine-tuning the SCL encoder on mined high-quality triplets and feeding the resulting representations into the prediction model.
From the experimental results, we can reach the following conclusions: 
\textbf{(1)} The annotation model captures incremental information beyond the baseline model; 
\textbf{(2)} By filtering and mining high-quality data based on the annotation model's results, and subsequently training the SCL Encoder, we achieve better performance than directly using the annotation model's scores.

\begin{table}[t]
    \centering
        \caption{Overall performance on the CTR prediction dataset.}
    \vspace{-0.3cm}
    \label{table6-1}
    \begin{tabular}{l|c|c}
    \toprule
  Method & GAUC & AUC\\
  \midrule
  Baseline~(include SCL) & - & - \\
  \midrule
  Add Annotation Score & +0.13\% & +0.08\% \\
  \midrule
  NMRL & \textbf{+0.22\%} & \textbf{+0.11\%} \\
  \bottomrule
    \end{tabular}
\vspace{-0.6cm}
\end{table}

\subsection{Online A/B Testing}
We integrate \textbf{N}ative \textbf{M}ultimodal \textbf{R}epresentation~(NMR) into Taobao's display advertising system in early 2026 and evaluate its online performance through a long-term A/B test. 
Results demonstrate that NMR \textbf{achieves a 1.5\% improvement in CTR~(Click-Through Rate) and a 0.5\% improvement in RPM~(Revenue Per Mille).}

\section{Related Work}
While discrete ID features play a pivotal role in recommendation models~\cite{1_1,1_2,1_3,1_4,1_5,1_6,1_7,DIN}, they suffer from inherent limitations, such as an inability to capture rich semantic information and vulnerability to cold-start scenarios~\citep{1_9, 1_10, 1_11, 1_12}. 
In contrast, multimodal representations directly leverage raw item content to capture deep semantic information, serving as a powerful complement to ID features. 
Most existing approaches adopt a two-stage paradigm to integrate multimodal features~\citep{2_4, 1_9, 2_5, 2_6, 2_7, 2_8, 2_9, simtier, notellm, notellm2, onerec, est, ts-rec}. Conversely, some recent studies have explored end-to-end training to better align multimodal representations with downstream tasks~\citep{qarm,lemur,em3}. 
However, a critical question remains unanswered: \textit{Does end-to-end training of the multimodal encoder still yield incremental gains when starting from a highly optimized, well-pretrained multimodal encoder?} 
Given the widespread adoption of the two-stage multimodal application paradigm in industry, resolving this question is of paramount importance. 
In this paper, we systematically investigate this problem and propose a novel, native multimodal representation learning framework.
\section{Conclusion}
In this paper, we focus on learning native multimodal representations for click-through rate~(CTR) prediction in e-commerce scenarios.
We demonstrate that E2EM fails because CTR behaviors are only partially attributable to multimodal semantics, while co-training with an ID-based CTR model still produces inconsistent encoder updates.
To address this, we propose \textbf{``Mine-Then-Train''}, which mines high-quality, semantically relevant triplets from CTR data with an annotation model and fine-tunes the SCL encoder on them.
This produces task-native multimodal representations and delivers a 1.5\% lift in CTR and a 0.5\% lift in RPM in Taobao’s online A/B tests.
Overall, our work provides a practical and effective paradigm for native multimodal representation learning in large-scale industrial recommendation systems.
\section*{GenAI Usage Disclosure}
Generative AI~(GenAI) tools were employed solely to polish the writing, improve grammatical correctness, and enhance the overall readability of this manuscript. 
All scientific conceptualizations, methodology designs, experimental interpretations, and core conclusions were entirely conceived and authored by the human authors, with no scientific content or findings generated by GenAI.

\bibliographystyle{ACM-Reference-Format}
\balance










\end{document}